\documentclass[12pt]{article} 
\usepackage[paper=a4paper, tmargin=15mm, bmargin=22mm, inner=15mm, outer=15mm]{geometry}
\usepackage{graphicx,floatflt,amssymb,rotate}
\usepackage{mathrsfs}
\usepackage{amssymb}
\usepackage{amsmath}
\usepackage{color}
\usepackage{graphicx}
\usepackage{subfig}
\usepackage{multirow}
\usepackage{nonfloat}
\usepackage{longtable}
\usepackage{footnote}
\usepackage{pstricks}
\usepackage[toc,page]{appendix}
\usepackage[mathscr]{euscript}
\usepackage{cite}
\usepackage[colorlinks=true,
linkcolor=red,
urlcolor=blue,
citecolor=blue]{hyperref}
\usepackage{color}

\usepackage{relsize}
\def\Babar{{\mbox{\slshape B\kern-0.1em{\smaller A}\kern-0.1em B\kern-0.1em{\smaller A\kern-0.2em R}}}}

\usepackage{enumitem}

\newcommand{\ba}{\begin{array}}
	\newcommand{\ea}{\end{array}}
\def\beq{\begin{equation}}
\def\eeq{\end{equation}}
\def\bea{\begin{eqnarray}}
\def\eea{\end{eqnarray}}
\def\nn{\nonumber}

\def\roughly#1{\mathrel{\raise.3ex\hbox
		{$#1$\kern-.75em\lower1ex\hbox{$\sim$}}}}

\def\sla#1{\raise.15ex\hbox{$/$}\kern-.57em #1}

\def\bd{B_d^0}

\def\order{\lower 1.8ex \hbox{\LARGE\~{}}}

\def\bd0tau{B\to D \tau\nu_{\tau}}

\def\be {\begin{equation}}
\def\ee {\end{equation}}

\usepackage[normalem]{ulem}

\definecolor{darkgreen}{cmyk}{1,0,1,0.4}

\definecolor{pink}{cmyk}{0.4,1,0.3,0}

\def\com2#1{\textcolor{red}{\it{#1}}}
\newcommand{\tcb}[1]{\textcolor{blue}{#1}}

\begin{document}

\begin{center}
	
	{\Large \bf $b \to c \ell \nu$ anomalies in light of extended scalar sectors}\\
	
	\vspace*{1cm}{\sf ~Aritra Biswas$^{a,}$\footnote{tpab2@iacs.res.in}, 
	                  ~Dilip Kumar Ghosh$^{a,}$\footnote{tpdkg@iacs.res.in}
                          ~Sunando Kumar Patra$^{b,}$\footnote{sunando.patra@gmail.com}
	                  ~Avirup Shaw$^{a,}$\footnote{avirup.cu@gmail.com}}\\
        \vspace{10pt} {\small } {\em $^a$ School of Physical Sciences, 
			Indian Association for the Cultivation of Science,\\
			2A $\&$ 2B, Raja S.C. Mullick Road, Kolkata 700032, India \\
	                $^b$ Department of Physics, Indian Institute of Technology, Kanpur-208016, India.}
\end{center}

	\begin{abstract} 
\noindent 
		Considering the recent experimental results on exclusive semileptonic $B$ meson decays showing sizable departure from their Standard Model prediction of lepton flavor universality and  keeping ongoing and proposed non-standard Higgs searches in mind, we explore the charged current flavor observables ($\mathcal{R}_{D^{(*)}}$, $\mathcal{R}_{J/\psi}$), among other $b\to c\ell \nu$ transitions, in the presence of a relevant scalar current effective new physics operator. We use $B_c$ lifetime and predicted bounds on the branching fraction of $B_c \to \tau \nu$ decay as constraints. We show the allowed parameter space in terms of the real and imaginary parts of the corresponding Wilson coefficients for such interactions. Under the light of obtained results, we study the prospect of two benchmark models, rendering the Wilson coefficients real (Georgi-Machacek (GM)) and complex (Leptoquark (LQ)) respectively. We show that constraints from $b\to c\ell \nu$ on GM parameters are consistent with other flavor constraints on the model, if we drop the \Babar~results. Including those disfavors the model by more than $3\sigma$. On the other hand, one benchmark LQ scenario, which gives rise to a single scalar current effective interaction, is still allowed within $68\%$ confidence level, albeit with a shrunk parameter space.
\vskip 5pt \noindent 
	\end{abstract}

\renewcommand{\thesection}{\Roman{section}}  
\setcounter{footnote}{0}  
\renewcommand{\thefootnote}{\arabic{footnote}}

	\section{Introduction}\label{intro}
	Over the last few years, one of the most engrossing puzzles in flavor physics has been the observed deviation of semileptonic $B$ decay observables from the corresponding Standard Model (SM) predictions. The reason for the excitement of the scientific community over these signals is because such deviations primarily hint towards the presence of lepton flavor universality violating (LFUV) new physics (NP). In this article we are particularly interested in the three following observables defined as:
	\begin{align}
	\nn\mathcal{R}_{D^{(*)}} &= \frac{\mathcal{B}(B\rightarrow D^{(*)}\,\tau\,\bar{\nu})}{ \mathcal{B}(B\rightarrow D^{(*)}\, e/\mu \, \bar{\nu})}\,,\\
	\mathcal{R}_{J/\psi} &= \frac{\mathcal{B}(B_c\rightarrow J/\psi \, \tau \, \bar{\nu})}{\mathcal{B}(B_c\rightarrow J/\psi \, \mu\, \bar{\nu})},
	\end{align}
	
	where $\mathcal{B}$ represents branching ratio. At the quark level, all these decays are induced by the $b\rightarrow c\, \ell \, \nu$ transitions  which appear at tree level in the SM. 
	
	A series of experiments have been done to measure these quantities by experimental collaborations such as \Babar~\cite{Lees:2012xj,Lees:2013uzd}, Belle \cite{Huschle:2015rga,Sato:2016svk,Hirose:2016wfn} and LHCb  \cite{Aaij:2015yra,Aaij:2017uff}, and most of them have reported the evidence for an excess in $\mathcal{R}_{D^{(*)}}$ from their SM predictions. 
	%
	%
	The current global averages for $\mathcal{R}_{D}$ and $\mathcal{R}_{D^*}$ deviate from the corresponding SM predictions by $\approx 2.2\sigma$ and $3\sigma$ respectively, with a combined deviation of $\approx 4\sigma$ \cite{average} with correlations taken into account. The ratio $\mathcal{R}_{J/\psi}$ has recently been measured by the LHCb collaboration \cite{Aaij:2017tyk}. This measurement also shows an excess of about $1.7\sigma$ from the central value range of the (rather imprecise) corresponding SM predictions $0.25-0.29$ \cite{Aaij:2017tyk, Ivanov:2005fd,Wen-Fei:2013uea, Dutta:2017xmj}.
	
	On a different note, the discovery of an SM-like Higgs boson at the Large Hadron Collider (LHC) \cite{Aad:2012tfa,Chatrchyan:2012xdj} in 2012 has triggered considerable interest in models with extended Higgs sector(s). The scientific community is searching for several non-standard (other than SM like) Higgs at the LHC and such searches will also be undertaken in other proposed future colliders like the International Linear Collider (ILC). Several theoretical models consisting of more than one Higgs with different charges have been proposed. These models can, in principle, resolve various shortcomings of the SM or can explain the observed deviation of several observables from the corresponding SM predictions. 

In view of the above observations, we explore the potential of an effective scalar type NP \cite{Crivellin:2012ye, Celis:2012dk, Crivellin:2015hha} for explaining such excesses in a model independent way \cite{Celis:2016azn} when the Wilson coefficient corresponding to such an operator is in general {\it complex} in nature. Thereafter, we use two benchmark models with such an operator for further illustration. The Georgi-Machacek (GM) model is one such model which contains a singly charged Higgs (with other scalars having different charges) that contributes to the $b\rightarrow c\, \ell \nu$ transitions at tree level. While several heavy flavor analyses \cite{Hartling:2014aga} have been performed for this model till date, semileptonic $B$ decays have never been studied in the GM model to the best of our knowledge. Results of our analysis show that the corresponding Wilson coefficient being real in nature, this model is disallowed upto ~$3\sigma$ by the $B_c\to\tau\nu$ constraint~\cite{Alonso:2016oyd}. On the other hand we consider a specific type of Leptoquark scenario which contains only one vector Leptoquark. In such a scenario we show that a complex Wilson coefficient is able to explain the available data on $b\rightarrow c\, \ell \nu$ transitions, albeit with the imaginary part being consistent with zero.
	
	
	The paper is organized in the following manner: in section~\ref{sec:obscons}, we introduce the observables and constraints to be analyzed and give the present status of their experimental determination. Section~\ref{sec:analys} contains the methods and results of our numerical analysis and in section~\ref{sec:models} we introduce two specific benchmark scenarios and discuss our results corresponding to those.
	
	\section{Observables and constraints:}\label{sec:obscons}
	\subsection{$\mathcal{R}_D$ and $\mathcal{R}_{D^*}$}
	
	The most general effective Hamiltonian describing the $b\to c\tau \nu$ transitions, with 
	all possible four-Fermi operators in the lowest dimension (with left-handed neutrinos)
	is given by \cite{Sakaki:2013bfa}
	\beq\label{eq1}
	{\cal H}_{eff} = \frac{4 G_F}{\sqrt{2}} V_{cb}\Big[( 1 + C^\ell_{V_1}) {\cal O}_{V_1} +
	C^\ell_{V_2} {\cal O}_{V_2} + C^\ell_{S_1} {\cal O}_{S_1}+ C^\ell_{S_2} {\cal O}_{S_2}+ C^\ell_{T}{\cal O}_{T}\Big],
	\eeq
	
	where the operator basis is defined as
	\bea
	{\cal O}_{V_1} &=& ({\bar c}_L \gamma^\mu b_L)({\bar \tau}_L \gamma_\mu \nu_{\tau L}) \nn, \\
	{\cal O}_{V_2} &=& ({\bar c}_R \gamma^\mu b_R)({\bar \tau}_L \gamma_\mu \nu_{\tau L}) \nn, \\
	{\cal O}_{S_1} &=& ({\bar c}_L  b_R)({\bar \tau}_R \nu_{\tau L}) \nn, \\
	{\cal O}_{S_2} &=& ({\bar c}_R b_L)({\bar \tau}_R \nu_{\tau L}) \nn, \\
	{\cal O}_{T}   &=& ({\bar c}_R \sigma^{\mu\nu} b_L)({\bar \tau}_R \sigma_{\mu\nu} \nu_{\tau L}),
	\label{eq2}
	\eea
	and the corresponding Wilson coefficients are given by $C_X (X=V_1,V_2,S_1,S_2,T)$. 
	We are interested in the new scalar interaction ${\cal O}_{S_1}$, and thus we turn all other Wilson coefficients to zero for this analysis. Following ref. \cite{Bhattacharya:2016zcw} and the references therein, differential decay rates for $B\rightarrow D^{(*)} \ell \nu_\ell$ (with $\ell = e$, $\mu$ or $\tau$ ) with this new interaction are given by:

	\begin{align}
		\frac{d\Gamma(\bar{B}\rightarrow D \ell \bar{\nu_\ell})}{dq^2} &= \frac{G^2_F |V_{cb}|^2}{96 \pi^3 m^2_B}q^2 p_D \left(1- \frac{m^2_\ell}{q^2}\right)^2\bigg[\left(1+ \frac{m^2_\ell}{2q^2}\right) H^{s 2}_{V, 0} \nn\\ 
		&+ \frac{3 m^2_\ell}{2 q^2} \left|1 + \frac{q^2}{m_{\ell}\left(m_b - m_c\right)} C^{\ell}_{S}\right|^2 H^{s 2}_{V, t} \bigg],\label{rd}\\
		\frac{d \Gamma(\bar{B}\rightarrow D^*\ell \bar{\nu_\ell})}{dq^2} &= \frac{G^2_F|V_{cb}|^2}{96(\pi)^3 m^2_B}q^2p_{D^*}\left(1-   \frac{m^2_\ell}{q^2}\right)^2 \bigg[(1+\frac{m^2_{\ell}}{2 q^2})\left(H^2_{V, +} + H^2_{V, -} + H^2_{V, 0}\right) \nn\\ 
		&+ \frac{3 m^2_{\ell}}{2 q^2} \left|1 + \frac{q^2}{m_{\ell}\left(m_b + m_c\right)} C^{\ell}_{S}\right|^2 H^2_{V, t}\bigg],\label{rdst}
	\end{align}
    
    where
	$p_{D^{(*)}}=\frac{\lambda^{1/2} (m^2_B,~m^2_{D^{(*)}},~q^2)}{2m_B}$ with $\lambda(a~,b,~c)=a^2+b^2+c^2-2(ab+bc+ca)$ and $q_\mu \equiv (p_B - p_X)_\mu$ is the momentum transfer (with $X = D$ or $D^{*}$). $H^s_{V, Y}(q^2)$ and $H_{V, Y}(q^2)$ are the helicity amplitudes for $\bar{B}\rightarrow D \ell \bar{\nu_\ell}$ and $\bar{B}\rightarrow D^* \ell \bar{\nu_\ell}$ transitions respectively (with $Y = \pm, ~0$ and $t$). These amplitudes can be  written down in terms of form factors which for the $B\rightarrow D^{(*)}$ transitions are taken from \cite{Fajfer:2012vx}. 
	We have dropped the subscript $1$ from $C^\ell_{S_1}$ as this is the only NP interaction we are considering and will follow this We follow the Caprini-Lellouch-Neubert (CLN) \cite{Caprini:1997mu} parametrization of the $B\to D^{(*)}$ form factors in this work. We have used both the fitted and predicted values of these parameters obtained in \cite{Jaiswal:2017rve}. In terms of the differential distributions of eqs.~\ref{rd} and \ref{rdst}, the ratios $\mathcal{R}(D^{(*)})$ 
	are defined as
	\beq\label{Rth}
	\mathcal{R}_{D^{(*)}} = \left[\int^{q^2_{max}}_{m^2_{\tau}} \frac{d\Gamma\left(\overline{B} \rightarrow D^{(*)}
		\tau \overline{\nu}\right)}{d q^2} d q^2\right]\left[\int^{q^2_{max}}_{m^2_{\ell}} 
	\frac{d\Gamma\left(\overline{B} \rightarrow D^{(*)} \ell \overline{\nu}\right)}{d q^2} d q^2\right]^{-1},
	\eeq
	with $q^2_{max}= (m_B - m_{D^{(*)}})^2$, and $\ell=e$ or $\mu$. 
	
	\subsection{$\mathcal{R}_{J/\psi}$}\label{sec:Rjpsi}
	Following eq.~\ref{eq1}, the differential decay rate for $\bar{B}\rightarrow J/\psi \ell \bar{\nu_\ell}$ (with $\ell = e$, $\mu$ or $\tau$ ) is given by:
	\bea
	\frac{d \Gamma(\bar{B_c}\rightarrow J/\psi\ell \bar{\nu_\ell})}{dq^2} &=& \frac{G^2_F|V_{cb}|^2}{96(\pi)^3 m^2_B}q^2p_{J/\psi}\left(1- \frac{m^2_\ell}{q^2}\right)^2\bigg[\left(1+\frac{m^2_{\ell}}{2 q^2}\right) \left(H^2_{J, +} + H^2_{J, -} + H^2_{J, 0}\right) \\ \nn
	&&+\frac{3 m^2_{\ell}}{2 q^2} (1 + \frac{q^2}{m_{\ell}\left(m_b + m_c\right)} C^{\ell}_{S})^2 H^2_{J, t}\bigg]\,,
	\label{jpsi}
	\eea
	where $p_{J/\psi}=\frac{\lambda^{1/2} (m^2_{B_c},~m^2_{J/\psi},~q^2)}{2m_{B_c}}$, with $\lambda(a,b,c)$ defined in the previous section. $H^s_{J, Y}(q^2)$ are the helicity amplitudes for the $\bar{B}\rightarrow J/\psi \ell \bar{\nu_\ell}$ transition(with $Y = \pm, ~0$). These amplitudes can be  written down in terms of form factors in a similar fashion as in the case of $\left(\overline{B} \rightarrow D^{(*)}	\ell \overline{\nu}\right)$ and a ratio can be defined by replacing the respective mesons in eq. \ref{Rth}. The theoretical predictions for the form factors and in turn the SM prediction of $\mathcal{R}_{J/\psi}$ is really imprecise till date and are yet to be tested against data. For this reason, the experimental group has used a $z$-expansion parametrization \cite{Bourrely:2008za} for the shared form factors for the signal and normalization modes and has determined them directly from the data. Those fit-results being unavailable, we need to choose from a plethora of available parametrizations in the literature, viz: the perturbative QCD (PQCD) approach~\cite{Wen-Fei:2013uea}, the constituent quark model~\cite{Anisimov:1998uk}, relativistic quark model~\cite{Ebert:2003cn}, non-relativistic quark model~\cite{Hernandez:2006gt}, QCD sum rules~\cite{Kiselev:2002vz}, relativistic constituent quark model~\cite{Ivanov:2006ni} and light-front covariant quark  model (LFCQ)~\cite{Wang:2008xt}.
	Choosing different parametrizations results in varying the central value of $\mathcal{R}_{J/\psi}$ within the range 0.25 - 0.29. We consider two parametrizations residing at  two far ends of this range, namely PQCD and LFCQ, to predict the SM value of $\mathcal{R}_{J/\psi}$,
	\bea
	\mathcal{R}^{\rm SM}_{J/\psi} &=& 0.249(42) ~~~~~{\rm LFCQ}\,,\nn\\ 
	&=& 0.289(28) ~~~~~{\rm PQCD}\,,
        \label{RJSM}
	\eea
	and perform our analyses with both sets of theoretical inputs. As is evident from eq.~\ref{RJSM}, the PQCD result is a little closer to the recent LHCb result, $\mathcal{R}^{\rm Exp.}_{J/\psi} = 0.71(17)_{st.}(18)_{sy.}$ \cite{Aaij:2017tyk}. Taking the uncertainties from different parameterizations into consideration, we see that the allowed theoretical range of $\mathcal{R}_{J/\psi}$ is larger than that considered in the LHCb analysis.
	
	\subsection{$P_{\tau}(D^*)$}
	Along with an update on $\mathcal{R}(D^*)$, ref.~\cite{Hirose:2016wfn} had announced the first ever measurement of the $\tau$ lepton polarization $P_{\tau}(D^*)$ in 2016. Though the result is imprecise and is consistent with SM, it is an observable we include in our analysis, as there is a considerable amount of correlation between this data-point and the $\mathcal{R}_{D^{(*)}}$ value measured in the same work. Studying further $\tau$ decays, $\tau$-polarization is defined as,
	\beq\label{ptaudef1}
	P_{\tau}(D^*) = \frac{\Gamma^{\lambda_{\tau}=1/2} - \Gamma^{\lambda_{\tau}= -1/2}}{\Gamma^{\lambda_{\tau}=1/2} + \Gamma^{\lambda_{\tau}= -1/2}}\,,{\rm where  ~~~} \Gamma^{\lambda_{\tau}=\pm 1/2} = \int_{m_{\tau}^2}^{q^2_{max}}\frac{d\Gamma^{\lambda_{\tau}=\pm 1/2}(\bar{B}\rightarrow D^*\tau \bar{\nu})}{dq^2}\,,
	\eeq
	$\lambda_{\tau}$ is the $\tau$ helicity, and $q^2_{max}= (m_B - m_{D^{*}})^2$. Following ref. \cite{Sakaki:2013bfa}, we can write,
		\begin{align}
		\nn \frac{d \Gamma^{\lambda_{\tau}= + 1/2} (\bar{B}\rightarrow D^*\tau \bar{\nu})}{dq^2} &= \frac{G^2_F|V_{cb}|^2}{96(\pi)^3 m^2_B}q^2p_{D^*}\left(1- \frac{m^2_\tau}{q^2}\right)^2 \frac{m^2_{\tau}}{2 q^2} \left[\left(H^2_{V, +} + H^2_{V, -} + H^2_{V, 0}\right) \right.\\
		&\left.+ 3 (1 + \frac{q^2}{m_{\tau}\left(m_b + m_c\right)} C^{\tau}_{S})^2 H^2_{V, t}\right]\,,\label{ptaudefp}\\
		\frac{d \Gamma^{\lambda_{\tau}= - 1/2} (\bar{B}\rightarrow D^*\tau \bar{\nu})}{dq^2} &= \frac{G^2_F|V_{cb}|^2}{96(\pi)^3 m^2_B}q^2p_{D^*}\left(1- \frac{m^2_\tau}{q^2}\right)^2 \left[\left(H^2_{V, +} + H^2_{V, -} + H^2_{V, 0}\right) \right]\,,\label{ptaudefm}
		\end{align}
	
	\subsection{$\mathcal{B}(B_c \to \tau \nu)$}\label{sec:bctn}
	Though $B_c$ mesons are produced copiously in LHCb, enabling precise determination of its lifetime, the decay $B_c \to \tau \nu$ remains out of the experimental reach for now \cite{Gouz:2002kk}. 
	In the presence of a new scalar operator, the branching fraction of $B_c \to \tau \nu$ can be expressed as \cite{Watanabe:2017mip,Alonso:2016oyd},
	\beq
	\mathcal{B}(B_c \to \tau \nu) = \tau_{B_c} \frac{m_{B_c} m^2_{\tau} f^2_{B_c} G^2_F \left|V_{cb}\right|^2}{8 \pi} \left(1 - \frac{m^2_{\tau}}{m^2_{B_c}}\right)^2\left|1 + \frac{m^2_{B_c}}{m_\tau (m_b + m_c)} C^{\tau}_S\right|^2\,,
	\label{Brbctn}
	\eeq
	where $f_{B_c} = 0.434(15)$GeV and $\tau_{B_c} = 0.507(9)$ps are the $B_c$ decay constant and the $B_c$ lifetime respectively. As the main contribution to $B_c$ lifetime should mainly be from $b$ and $c$ decays in the $B_c$ meson,  accounting for the maximum possible errors in the calculation, only $\lesssim 30\%$ of the measured experimental width ($\Gamma_{B_c} = 1 / \tau_{B_c}$) can be explained by (semi)tauonic modes, even after including any NP effects that could potentially explain the $\mathcal{R}(D^{(*)})$ excess \cite{Alonso:2016oyd}.
	
	An even stronger upper bound of $\lesssim 10\%$ is obtained from LEP data taken at $Z$-peak \cite{Akeroyd:2017mhr} and there is a prospect of making the bound even tighter by doing an analysis with the full L3 data \cite{Acciarri:1996bv}.
	In our analysis, we have used two constraints. The relaxed one comes due to considering $\mathcal{B}(B_c \to \tau \nu) \lesssim 30\%$ and the tighter one by considering $\mathcal{B}(B_c \to \tau \nu) \lesssim 10\%$.

	\section{Analysis}\label{sec:analys}
	
	\subsection{Present Status}\label{status}
	
	The current experimental status with appropriate correlations as well as the SM values for $\mathcal{R}_{D^{(*)}}$ and $P_{\tau}(D^*)$ are displayed in Table~\ref{tab:RDRDsPtau}. Though there are numerous SM predictions for these quantities quoted in literature \cite{Na:2015kha,Kamenik:2008tj,Lattice:2015rga,Bigi:2016mdz,Bigi:2017jbd,Bernlochner:2017jka,Tanaka:2012nw,Ivanov:2016qtw}, we follow the results from the analysis in ref.~\cite{Jaiswal:2017rve} and calculate the SM results ourselves. This gives us the opportunity to make use of the theoretical correlations between $\mathcal{R}_{D^{(*)}}$ and $P_{\tau}(D^*)$ in our analysis (listed in Table~\ref{tab:RDDPtaucor}).
	
	Recently, Belle has updated their data with semileptonic $B$-tag (`Belle (2019)' in Table \ref{tab:RDRDsPtau}), which now supersedes the earlier result (`Belle (2016)-I'). To showcase the difference between these results and the effect of the newer one, we perform our analysis in two separate sets. For the case involving the ``old'' dataset, we consider the Belle (2016)-I data and drop the Belle (2019) data. For this case, we have a total of ten data points (two $\mathcal{R}(D)$, six $\mathcal{R}(D^*)$, one $\mathcal{R}_{J/\psi}$, and one $P_{\tau}(D^*)$). The ``new'' dataset involves the Belle (2019) data with the Belle (2016)-I dropped, resulting in a total of 11 data points. It is to be noted that apart from Belle 2015, Belle (2019) and the LHCb (2017) result, the rest are consistent with a sizable deviation from the SM. In addition to verifying whether scalar NP operators (which are already known to be heavily constrained, but not ruled out, from the $\mathcal{R}(D^*)$ data) are still a viable candidate for explaining lepton flavor universality violating charged current anomalies, the effect of the addition of the $\mathcal{R}_{J/\psi}$ data on the single scalar parameter case might also be of interest. Hence, we refrain from using the global averages.

\begin{savenotes}
\begin{table*}[ht!]
		\centering
                     \resizebox{17cm}{!}{
				\begin{tabular}{|c|c|c|c|c|}\hline
					& $\mathcal{R}(D)$  & $\mathcal{R}(D^*)$  &	$\to$ Correlation	& $P_{\tau}(D^*)$ \\
					\hline
					SM			& $0.304(3)$		& $0.259(6)$		  &						&  $-0.491(25)$\\
					\hline
					\Babar~   	& $0.440(58)_{st.}(42)_{sy.}$ 	& $0.332(24)_{st.}(18)_{sy.}$ & $-0.31$\cite{Lees:2013uzd} & \\
					Belle (2015)    & $0.375(64)_{st.}(26)_{sy.}$ 	& $0.293(38)_{st.}(15)_{sy.}$ & $-0.50$\cite{Huschle:2015rga}&\\
					Belle (2016)-I  & -			 	& $0.302(30)_{st.}(11)_{sy.}$ \cite{Abdesselam:2016cgx}\footnote{This result is included in the ``old dataset'' case exhibited in Table.~\ref{tab:res1} and is superseded by the Belle (2019) data which goes into the analysis for the ``new dataset'' also shown in the same table.} & & \\
					Belle (2016)-II &    & $0.270(35)_{st.}~^{+ 0.028}_{-0.025}$ & 0.33\footnote{This correlation is between $\mathcal{R}(D^*)$ and $P_{\tau}(D^*)$. Stat. corr. = 0.29 and syst. corr. = 0.55.} & $ -0.38(51)_{st.}~^{+0.21}_{-0.16}$ \cite{Hirose:2016wfn}\\
					Belle (2019) &$0.307(37)_{st.}(16)_{sy.}$ & $0.283(18)_{st.}(14)_{sy.}$ & $-0.51$\cite{Abdesselam:2019dgh} & \\
					LHCb (2015) & - 				& $0.336(27)_{st.}(30)_{sy.}$ \cite{Aaij:2015yra} & & \\
					LHCb (2017) & - 				& $0.286(19)_{st.}(25)_{sy.}(21)$\footnote{This uncertainty originates from the uncertainties on $\mathcal{B}(B^0\to D^{*-}\pi^+\pi^-\pi^+)$ and $\mathcal{B}(B^0\to D^{*-}\mu^+\nu_{\mu})$.} \cite{Aaij:2017deq} & & \\
					\hline
					World Avg.	& $0.340(27)_{st.}(13)_{sy.}$	& $0.295(11)_{st.}(8)_{sy.}$ & $-0.38$ \cite{average} & \\ \hline
				\end{tabular}}
				\caption{Present status (both theoretical and experimental) of $\mathcal{R}(D)$, $\mathcal{R}(D^*)$ and $P_{\tau}(D^*)$. First uncertainty is statistical and the second one is systematic. The first row lists the SM calculation obtained in this paper.} 
				\label{tab:RDRDsPtau}
	\end{table*}
  \end{savenotes}
	
	\begin{table*}[ht!]
		\centering
                     \resizebox{6cm}{!}{
				\begin{tabular}{|c|c|c|c|}\hline
					& $\mathcal{R}(D)$  & $\mathcal{R}(D^*)$  & $P_{\tau}(D^*)$ \\
					\hline
					$\mathcal{R}(D)$  	& 1. 				& 0.118 			  & -0.023 \\
					$\mathcal{R}(D^*)$  & 					& 1. 				  & 0.617 \\ 
					$P_{\tau}(D^*)$ 	&					& 					  & 1. \\ \hline
				\end{tabular}}
				\caption{Theoretical correlations between the SM values of observables listed in the first row of Table \ref{tab:RDRDsPtau}.} 
				\label{tab:RDDPtaucor}
			\end{table*}
\vspace*{-1cm}
	
	\subsection{Methodology}

	
	We take the dimensionless Wilson coefficient $C^{\tau}_{S}$ to be complex in this part of the analysis.
	All subsequent parameter spaces for $C^{\tau}_{S}$ in this work are obtained by optimizing a $\chi^2$ statistic using \textit{Mathematica}\textsuperscript{\textcopyright} in the form of a package \cite{OptEx}. While the parameter confidence levels (CLs) are  obtained without any external constraints, the constraint due to the branching ratio for the $B_c\to\tau\nu$ mode is overlapped on the obtained CLs. Further discussion regarding this constraint can be found following the relevant figures related to our analysis. The $\chi^2$ statistic is defined as:
	\beq\label{chidef}
	\chi^2 (C^{\tau}_{S}) = \sum^{{\rm data}}_{i,j = 1} \left(Obs^{{exp}}_i - Obs^{th}_i(C^{\tau}_{S})\right)
	\left(V^{stat} + V^{syst} \right)^{-1}_{i j} \left(Obs^{{exp}}_j - Obs^{th}_j(C^{\tau}_{S})\right)
	+ \chi^2_{Nuis.}\,.
	\eeq
	Here, $Obs^{th}_k(C^{\tau}_{S})$ is given by eqs.~\ref{Rth}, \ref{jpsi}, and \ref{ptaudef1} as applicable and $Obs^{{exp}}_k$ is the central value of the k$^{th}$ experimental result. In constructing the statistical (systematic) covariance matrices $V^{stat (syst)}$, we have taken separate correlations, wherever available. The nuisance parameters (Table~\ref{tab:nuisance}) occurring in the theoretical expressions are tuned in to the fit using a term
	\beq\label{chifit}
	\chi^2_{Nuis.} = \sum^{{\rm theory}}_{i,j = 1} \left({\rm param}_i - {\rm value}_i\right)~\left(V^{Nuis}\right)^{-1}_{i j}\\
	\left({\rm param}_j - {\rm value}_j\right)\,.
	\eeq
	For each of the cases subject to the ``old'' and  ``new'' data-sets as discussed in section~\ref{status}, we perform the fits in two separate stages. First, we use the experimental results for $\mathcal{R}_{D^{(*)}}$ and $P_{\tau}(D^*)$ listed in Table~\ref{tab:RDRDsPtau} to fit $C^{\tau}_{S}$. In the second stage, we redo the fits with $\mathcal{R}_{J/\psi}$ included in the $\chi^2$ of eq.~\ref{chidef}. The $\chi^2$ will now, in addition to the uncertainties of the parameters in Table~\ref{tab:nuisance}, contain the uncertainties corresponding $B_c\rightarrow J/\psi$ form factors. Following the discussion in sec. \ref{sec:Rjpsi}, we do two sets of fits in this stage, with two different sets of form factor parametrization, namely, LFCQ and PQCD.
	After each fit, we determine the quality of it in terms of the $p$-value obtained. This gives a direct and better quantifiable estimate of the quality of our fits than just the $\chi^2_{min}$ values, as the degrees of freedom (DoF) vary for different fits. Finally, we add the constraints mentioned in sec. \ref{sec:bctn} to our analysis and obtain the allowed parameter space. 

\begin{table*}[ht!]
		\centering
                     \resizebox{10cm}{!}{
				\begin{tabular}{|c|c|c|c|c|c|c|}\hline
					Parameters 	& Value		&\multicolumn{5}{|c|}{Correlation}\\ 
					\hline 
					$\rho_D^2$ 	& 1.138(23) 	& 	1. 	& 0.15 	& -0.01 & -0.07 & 0 \\
					$\rho_{D^*}^2$ & 1.251(113)	&  		& 1. 	& 0.08 	& -0.80 & 0 \\
					$R_1(1)$ 	& 1.370(36) 	&  		& 		& 1. 	& -0.08 & 0 \\
					$R_2(1)$ 	& 0.888(65) 	&  		& 		& 		& 1. 	& 0 \\
					$R_0(1)$ 	& 1.196(102) 	&  		&  		&  		&  		& 1 \\
					\hline
					$m_B$		& 5.27962(15) GeV	& & & & & \\
					$m_{D^*}$	& 2.01026(5) GeV	& & & & & \\
					$m_W$ 		& 80.385(15) GeV	& & & & & \\
					$m_W$ 		& 80.385(15) GeV	& & & & & \\
					$m_c$ 		& 1.28(3) GeV	& & & & & \\
					$m_b$ 		& $4.18^{+0.04}_{-0.03}$ GeV	& & & & & \\
					$m_{\tau}$	& 1.77682(16) GeV	& & & & & \\ \hline
				\end{tabular}}
				\caption{Nuisance inputs in the theory expressions. Only those form factor parameters which appear in $\mathcal{R}_{D^{(*)}}$ and $P_{\tau}(D^*)$ are shown here. These are obtained from the analysis in ref. \cite{Jaiswal:2017rve}.} 
				\label{tab:nuisance}
	\end{table*}

	
	\subsection{Results and discussions}\label{sec:modindres}
	
	Gist of the results obtained in our numerical analysis is encapsulated in Table~\ref{tab:res1}. For both the cases discussed in the previous section (with or without the $\mathcal{R}_{J/\psi}$ result), we exhibit the $68.27\%$ and $95.45\%$ CLs for the ``new'' and ``old'' datasets as discussed in section~\ref{status}. Excluding $\mathcal{R}_{J/\psi}$, we find that the the old dataset yields a much  better fit than the new one, as can clearly be inferred from the corresponding p-values in table~\ref{tab:res1}. The reason can be tracked back to the fact that the Belle (2016)-I data point, which is included in the old dataset, is farther away from the SM in comparison to the Belle (2019) one. All other members of these datasets being the same, the new dataset hence yields a worse fit for NP in comparison to the old one. The same pattern can be observed even after the inclusion of $\mathcal{R}_{J/\psi}$. We perform these fits with two different sets of form factor parametrization as mentioned in sec. \ref{sec:Rjpsi}. While the parameter space remains more or less the same in comparison to the fits without $\mathcal{R}_{J/\psi}$, the new dataset consistently yields a poorer fit in comparison to the old one. 

\begin{table*}[ht!]
		\centering
                       \resizebox{11cm}{!}{
				\begin{tabular}{|c|c|c|c|c|c|}\hline
					\multicolumn{2}{|c|}{Dataset}              & $\chi^2_{min}$ & $p$-value & \multicolumn{2}{|c|}{Fit Results}  \\ 
					\cline{5-6}
					\multicolumn{2}{|c|}{}                     &  /DoF          &    (\%)      & Re($C^{\tau}_{S}$)                        & Im($C^{\tau}_{S}$)  \\ 
					\hline
					Old     & without $\mathcal{R}_{J/\psi}$   & 9.23/7         & 23.65     &	$0.227^{+0.165}_{-0.058}$	    & $0.0\pm0.51$	       	  \\ 
					\cline{2-6}
					        & with $\mathcal{R}_{J/\psi}$ PQCD & 11.95/8        & 15.34     &	$0.228^{+0.159}_{-0.058}$	    & $0.0\pm0.5$	       	  \\ 
					\cline{2-6}
					        & with $\mathcal{R}_{J/\psi}$ LFCQ & 12.51/8        & 12.98     &	$0.227^{+0.161}_{-0.058}$	    & $0.0\pm0.5$	       	  \\ 
					\hline
					New     & without $\mathcal{R}_{J/\psi}$   & 14.16/8	    &  7.77     &       $0.111^{+0.152}_{-0.043}$	    &  $0.0\pm0.45$               \\ 
					\cline{2-6}
					        & with $\mathcal{R}_{J/\psi}$ PQCD & 16.95/9	    & 4.95      &	$0.111^{+0.146}_{-0.043}$	    & $0.0\pm0.45$	          \\ 
					\cline{2-6}
					        & with $\mathcal{R}_{J/\psi}$ LFCQ & 17.49/9	    & 4.16      &	$0.111^{+0.148}_{-0.043}$	    & $0.0\pm0.45$	          \\ 
					\hline
				\end{tabular}}
				\caption{Results of fits for the with and without $\mathcal{R}_{J/\psi}$ cases involving both the old and new datasets. Detailed analysis is in section~\ref{sec:modindres}.} 
				\label{tab:res1}
	\end{table*}
	
	\begin{figure*}[ht!]
		\centering
		\subfloat[without $\mathcal{R}_{J/\psi}$]{\includegraphics[width=0.5\linewidth]{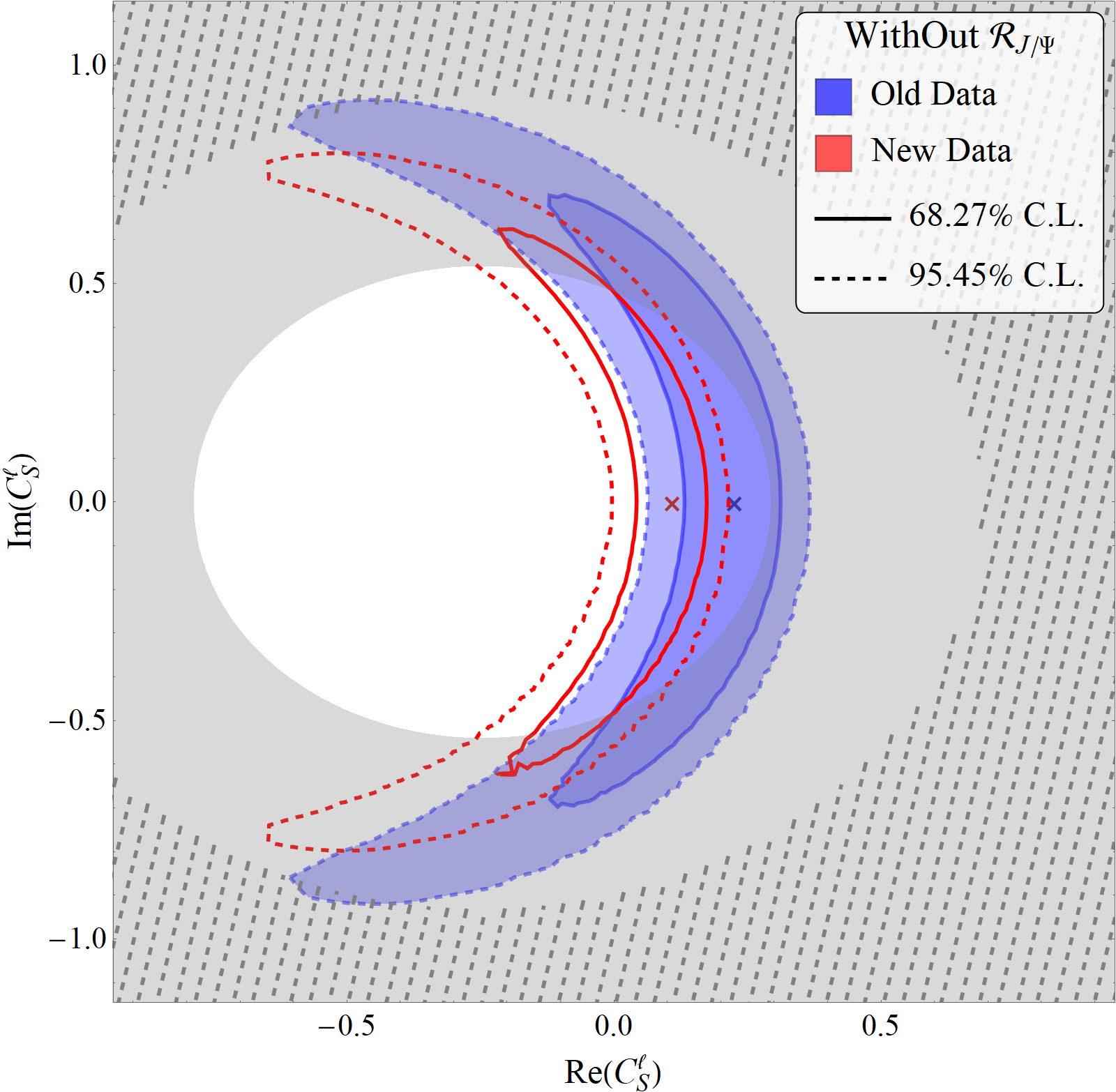}}
		\subfloat[with $\mathcal{R}_{J/\psi}$ (PQCD)]{\includegraphics[width=0.5\linewidth]{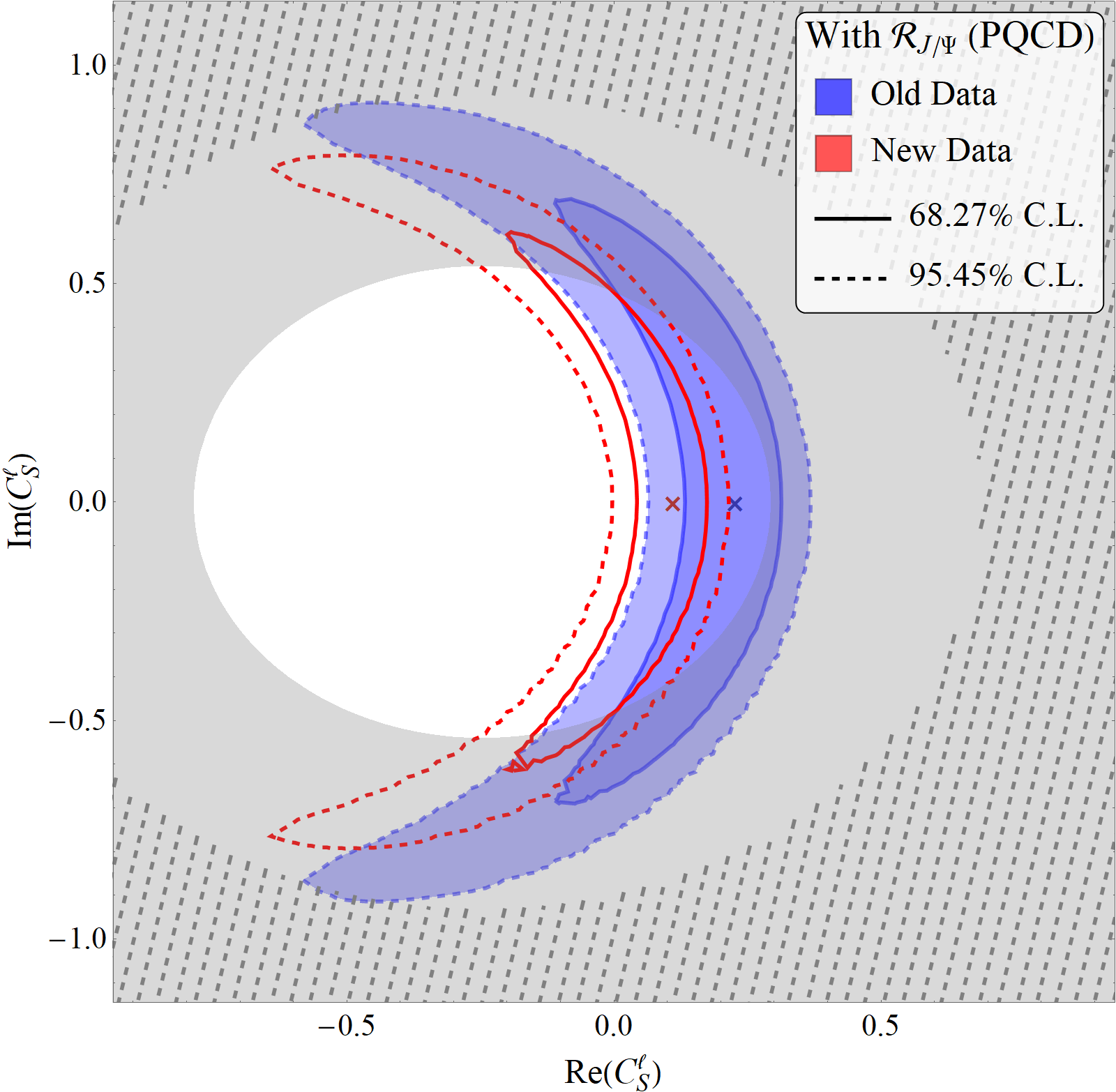}}\\
		\subfloat[with $\mathcal{R}_{J/\psi}$ (LFCQ)]{\includegraphics[width=0.5\linewidth]{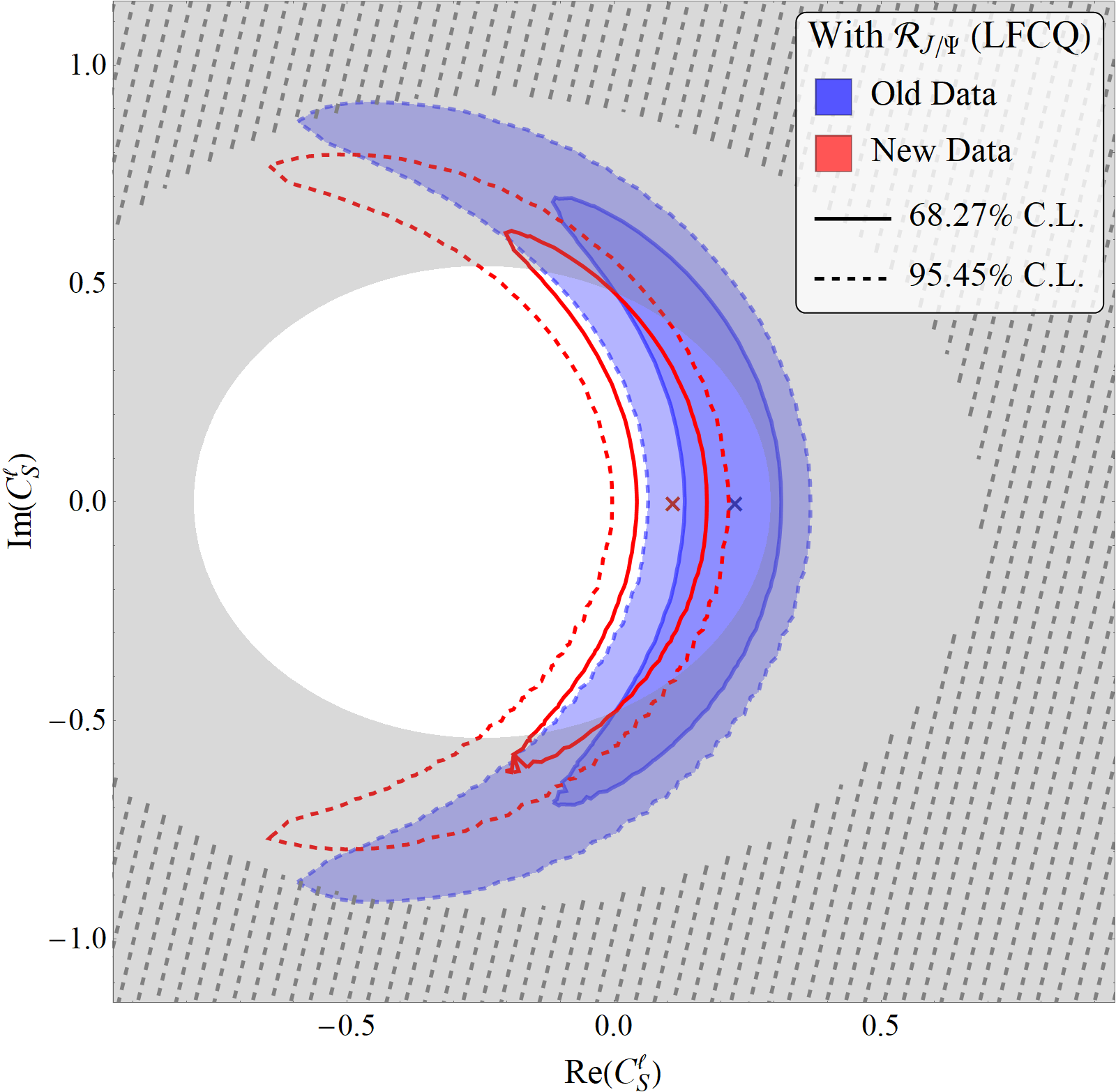}}
		\caption{Fit results in terms of the fixed $\Delta \chi^2$ contours representing $68.27\%$ (solid) and $95.45\%$ (dot-dashed) CLs respectively, in the Re($C^{\tau}_{S}$) and Im($C^{\tau}_{S}$) plane (a) without $\mathcal{R}_{J/\psi}$, (b) with $\mathcal{R}_{J/\psi}$ parametrized by PQCD form factors and (c) with $\mathcal{R}_{J/\psi}$ involving the LFCQ form factor parametrization. The blue filled region corresponds to the old dataset, while the red transparent one stands for the new dataset. The diagonally hatched region is ruled out by the $\mathcal{B}(B_c \to \tau \nu) \lesssim 30\%$ constraint and the gray-shaded region is disallowed by the constraint $\mathcal{B}(B_c \to \tau \nu) \lesssim 10\%$.}
		\label{fig:rcHicH}
	\end{figure*}
	An interesting aspect of these results is that the $\mathcal{R}_{J/\psi}$, while in tension with the SM prediction, is in tension with other data-points as well, with the pull in the opposite direction. In fact, for the new dataset, the inclusion of $\mathcal{R}_{J/\psi}$ is unable to yield an acceptable fit (p-value being less than 5\%. In other words, the obtained best-fit values for all datasets without the $\mathcal{R}_{J/\psi}$ as a data-point, will give a prediction of $\mathcal{R}_{J/\psi}$ that lies in tension with the experimentally measured value. This finding is consistent with the conclusion of a recent work \cite{Alok:2017qsi} that the central value of experimentally measured $\mathcal{R}_{J/\psi}$ is too high to be explained by any NP. As the phase space ratio of the quark level transitions are same for $\mathcal{R}_{D^{(*)}}$ and $\mathcal{R}_{J/\psi}$, the amount of NP needed to explain the deviation in $\mathcal{R}_{D^{(*)}}$ is insufficient to reach anywhere near the high central value of the measured $\mathcal{R}_{J/\psi}$. As the PQCD prediction lies closer to the experimental result of $\mathcal{R}_{J/\psi}$ than that of the LFCQ one, the fits for PQCD are, in general, slightly better. Another feature of these fits is that correlations between the fit parameters are consistent with zero up to four decimal places and we chose not to show them in the table.
	
	Fig.~\ref{fig:rcHicH} shows the two dimensional parameter spaces for the cases involving and excluding the $\mathcal{R}_{J/\psi}$ data for both the old and new datasets.  Though the addition of $\mathcal{R}_{J/\psi}$ changes the quality of fits, the parameter space hardly changes. We have shown the allowed parameter space as regions bounded by fixed $\Delta\chi^2$ contours. The solid contours depict the 68.27\% CLs, while the dashed ones depict the 95.45\% CLs. The blue contours represent the old dataset, while the red ones stand for the new one. As an illustrative case, for a fit with $\mathcal{R}_{J/\psi}$ (form factors from PQCD) and old data, $\chi^2_{\rm SM} $= $25.80$, which puts the SM $\sim 3.3\sigma$ away from the best-fit point.
	
	Following the arguments from sec.~\ref{sec:bctn}, we have chosen to show two illustrative bounds coming from the $\mathcal{B}(B_c \to \tau \nu)$, the relaxed one being $= 30\%$ and the aggressive one as $= 10\%$. These are depicted in the plots as diagonally hatched and smooth gray circular shells respectively. We can see that a considerable amount of parameter space is allowed even after invoking these constraints and the best-fit points lie well within the allowed region for all the cases discussed above. However, the p-values corresponding to the with $\mathcal{R}_{J/\psi}$ case for the new dataset are less than $5\%$ significant (see table~\ref{tab:res1}) and the resulting fit is hence not acceptable from a frequentest point of view. 
	
	\section{Interpreting the fit results: A few examples}\label{sec:models}
	
	Within the scope of this article we have so far explored the prospect for a scalar type NP in explaining the present set of charged current anomalies that have recently been experimentally measured. In this section, we attempt to exhibit the effects of our fit results on two general class of models, where the scalar operator is preceded by ($a$) a real and ($b$) a complex Wilson coefficient.
	We choose two NP scenarios to illustrate these cases. As an example of the former, we choose the GM model, while we choose a specific type of Leptoquark model to illustrate the latter. In the following, model specific analyses are provided after a short description of each one.
	
	\subsection{The Georgi-Machacek model}\label{3}
	\subsubsection{Description}
	
	Here we briefly discuss the Georgi-Machacek (GM) model \cite{Georgi:1985nv} whose scalar sector comprises of the usual complex doublet $(\phi^+ ~~\phi^0)$ of hypercharge $Y = 1$, a real triplet $(\xi^+~~\xi^0~~\xi^-)$ with $Y = 0$, and  a complex triplet $(\chi^{++}~~\chi^+~~\chi^0)$ with $Y=2$. In order to achieve the global $SU(2)_L \times SU(2)_R$ symmetry explicitly, one can express the doublet in the form of a bi-doublet $\varphi$ and combine the triplets to form a bi-triplet $X$ \cite{Hartling:2014zca, Hartling:2014aga}:
	\begin{eqnarray}
	\nn\varphi &=& \left( \begin{array}{cc}
	\phi^{0*} &\phi^+  \\
	-\phi^{+*} & \phi^0  \end{array} \right)~{\rm and}~ 
	X=
	\left(
	\begin{array}{ccc}
	\chi^{0*} & \xi^+ & \chi^{++} \\
	-\chi^{+*} & \xi^{0} & \chi^+ \\
	\chi^{++*} & -\xi^{+*} & \chi^0  
	\end{array}
	\right).\\
	\label{eq:PX}
	\end{eqnarray}
	The most general scalar potential involving these fields that conserves custodial $SU(2)$ and respects the gauge invariance is given by \cite{Hartling:2014zca, Hartling:2014aga}
		\begin{eqnarray}
		V(\varphi,X) &= & \frac{\mu_2^2}{2}  \text{Tr}(\varphi^\dagger \varphi) 
		+  \frac{\mu_3^2}{2}  \text{Tr}(X^\dagger X)  
		+ {\bar{\lambda}}_{1} [\text{Tr}(\varphi^\dagger \varphi)]^2  
		+ {\bar{\lambda}}_{2} \text{Tr}(\varphi^\dagger \varphi) \text{Tr}(X^\dagger X)   \nonumber \\
		& & + {\bar{\lambda}}_{3} \text{Tr}(X^\dagger X X^\dagger X)  
		+ {\bar{\lambda}}_{4} [\text{Tr}(X^\dagger X)]^2 
		- {\bar{\lambda}}_{5} \text{Tr}( \varphi^\dagger \tau^a \varphi \tau^b) \text{Tr}( X^\dagger t^a X t^b) 
		\nonumber \\
		& & - M_1 \text{Tr}(\varphi^\dagger \tau^a \varphi \tau^b)(U X U^\dagger)_{ab}  
		-  M_2 \text{Tr}(X^\dagger t^a X t^b)(U X U^\dagger)_{ab}\;,
		\label{eq:potential}
		\end{eqnarray}
	where ${\bar{\lambda}}_{k}$ ($k=1,\ldots 5$) are real dimensionless coupling constants; $\mu_2$, $\mu_3$, $M_1$ and $M_2$ are the real mass parameters. Here $\tau^a = \sigma^a/2$ ($\sigma^a$ being the Pauli matrices) are the $SU(2)$ generators for the doublet representation and the generators for the triplet representation are
	\begin{align}
	\nn	t^1= \frac{1}{\sqrt{2}} \left( \begin{array}{ccc}
	0 & 1  & 0  \\
	1 & 0  & 1  \\
	0 & 1  & 0 \end{array} \right), \quad  
	t^2= \frac{1}{\sqrt{2}} \left( \begin{array}{ccc}
	0 & -i  & 0  \\
	i & 0  & -i  \\
	0 & i  & 0 \end{array} \right), \quad
        t^3= \left( \begin{array}{ccc}
	1 & 0  & 0  \\
	0 & 0  & 0  \\
	0 & 0 & -1 \end{array} \right),
	\end{align}
	
	where the matrix $U$, given in ref.~\cite{Aoki:2007ah}, rotates $X$ into the Cartesian basis.

	To ensure the conservation of custodial $SU(2)$, the neutral components of the triplet fields $\chi^0$ and $\xi^0$ must obtain the same vacuum expectation value (VEV). These neutral fields can be decomposed around their VEVs as \cite{Hartling:2014zca, Hartling:2014aga},
	\begin{align}
	\nn\phi^0 \to \frac{v_{\phi}}{\sqrt{2}} + \frac{\phi^{0,r} + i \phi^{0,i}}{\sqrt{2}},
	\quad
	\chi^0 \to v_{\chi} + \frac{\chi^{0,r} + i \chi^{0,i}}{\sqrt{2}},
        \quad 
        \xi^0 \to v_{\chi} + \xi^0.
	\end{align}
	After spontaneous symmetry breaking, one obtains the $W$ and $Z$ boson masses which further relate the doublet and triplet VEV's as \cite{Hartling:2014zca, Hartling:2014aga}
	\begin{equation}
	v_{\phi}^2 + 8 v_{\chi}^2 \equiv v^2 = (246~{\rm GeV})^2,
	\label{eq:vevrelation}
	\end{equation}
	where $v$ is the electroweak VEV.
	
	Under the custodial $SU(2)$ symmetry, the physical fields are organized by their transformation properties and can be considered as a quintuplet ($H^{++}_5, H^{+}_5, H^{0}_5, H^{-}_5, H^{--}_5$) and a triplet ($ H^{+}_3, H^{0}_3, H^{-}_3$). The masses are degenerate in each custodial multiplet at the tree level. There are two more singlets ($h, H$) and one of them can be considered as the SM-like Higgs. Along with the physical scalar fields, one has three Goldstone bosons ($G^+, G^-, G^0$).
	The presence of several non-standard scalars makes this model phenomenologically very attractive. One can find the phenomenology of different flavor observables in \cite{Hartling:2014aga}. Some collider studies in this scenario have been done in \cite{Chiang:2012cn, Chiang:2014bia, Chiang:2015rva, Chiang:2015amq, Sun:2017mue}. 

	Since this model is custodially protected, the $\rho$ parameter is equal to unity at tree level. As a result, the triplet VEV in this model is much loosely constrained in comparison to other models with extended Higgs sector (e.g. the type-II seesaw model~\cite{Arhrib:2011uy}). 
	
	\subsubsection{Analysis}
	
	Among the several non-standard scalars of the GM model, $H^{\pm}_3$ contributes to the $b\to c\ell \nu$ transition. Hence, we can have extra scalar type interaction, which we are searching for, to explain the anomaly of our interest in this article. In this model, the expression for $C^{\ell}_{S}$ is given by:
	\begin{equation}
	C^{\ell}_{S}=-C_H ~m_b ~m_{\ell} =-\frac{\tan^2\theta_H}{m^2_{H^\pm_3}} m_b ~m_{\ell}.
	\label{csl2}
	\end{equation}
	Here, $\theta_H$ is the mixing angle and is given by \cite{Hartling:2014zca, Hartling:2014aga}
	\begin{equation}
	\tan\theta_H = \frac{2\sqrt{2}\,v_\chi}{v_\phi}~,
	\end{equation}
	and ${m_{H^\pm_3}}$ is the mass of $H^\pm_3$. We can take the lower mass bound on this singly charged scalar obtained from the direct search  at the LEP II as 78 GeV \cite{Searches:2001ac}.
	\begin{figure*}[hbt]
		\centering
		\subfloat[without $\mathcal{R}_{J/\psi}$]{\includegraphics[width=0.5\linewidth]{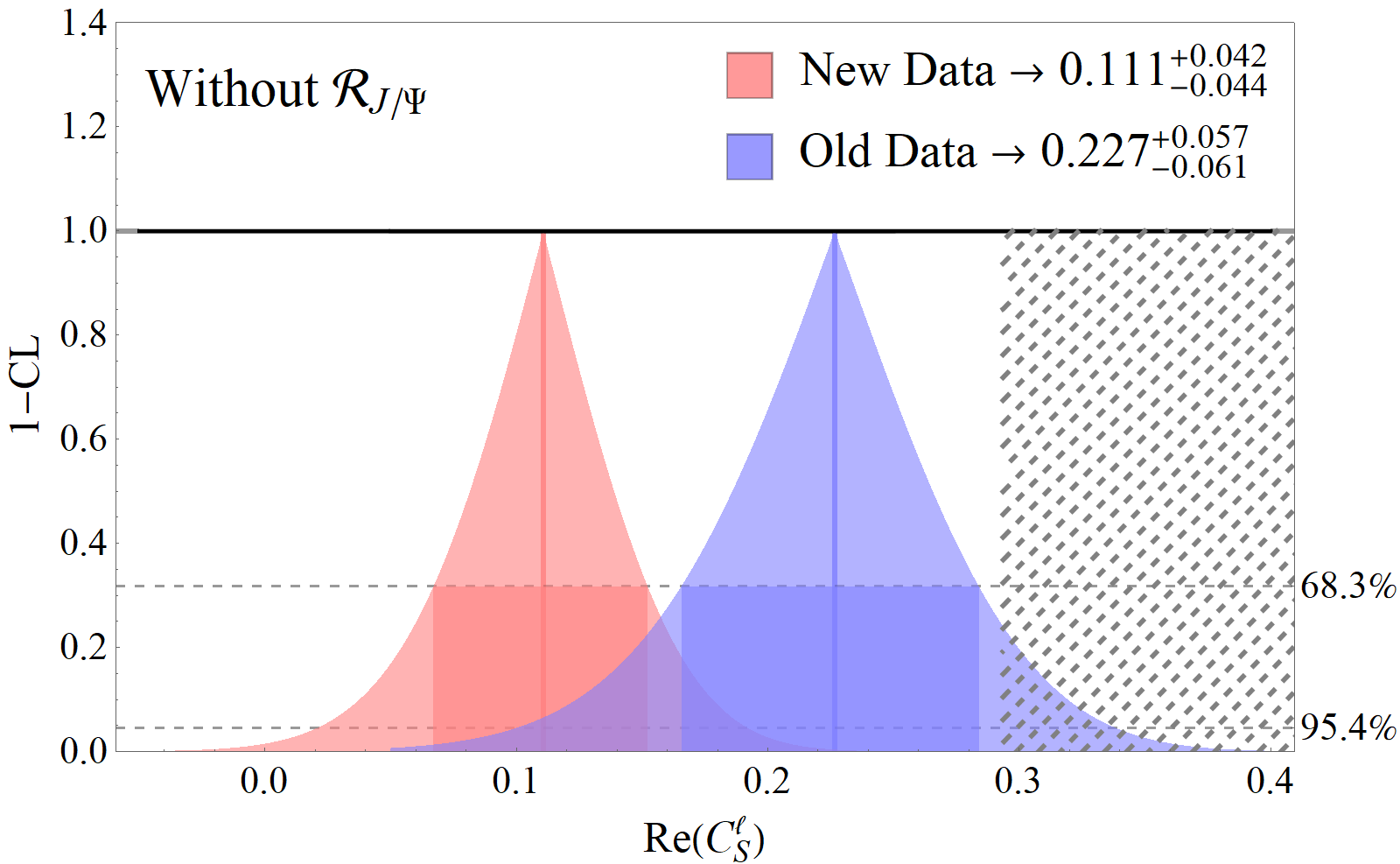}}
		\subfloat[PQCD]{\includegraphics[width=0.5\linewidth]{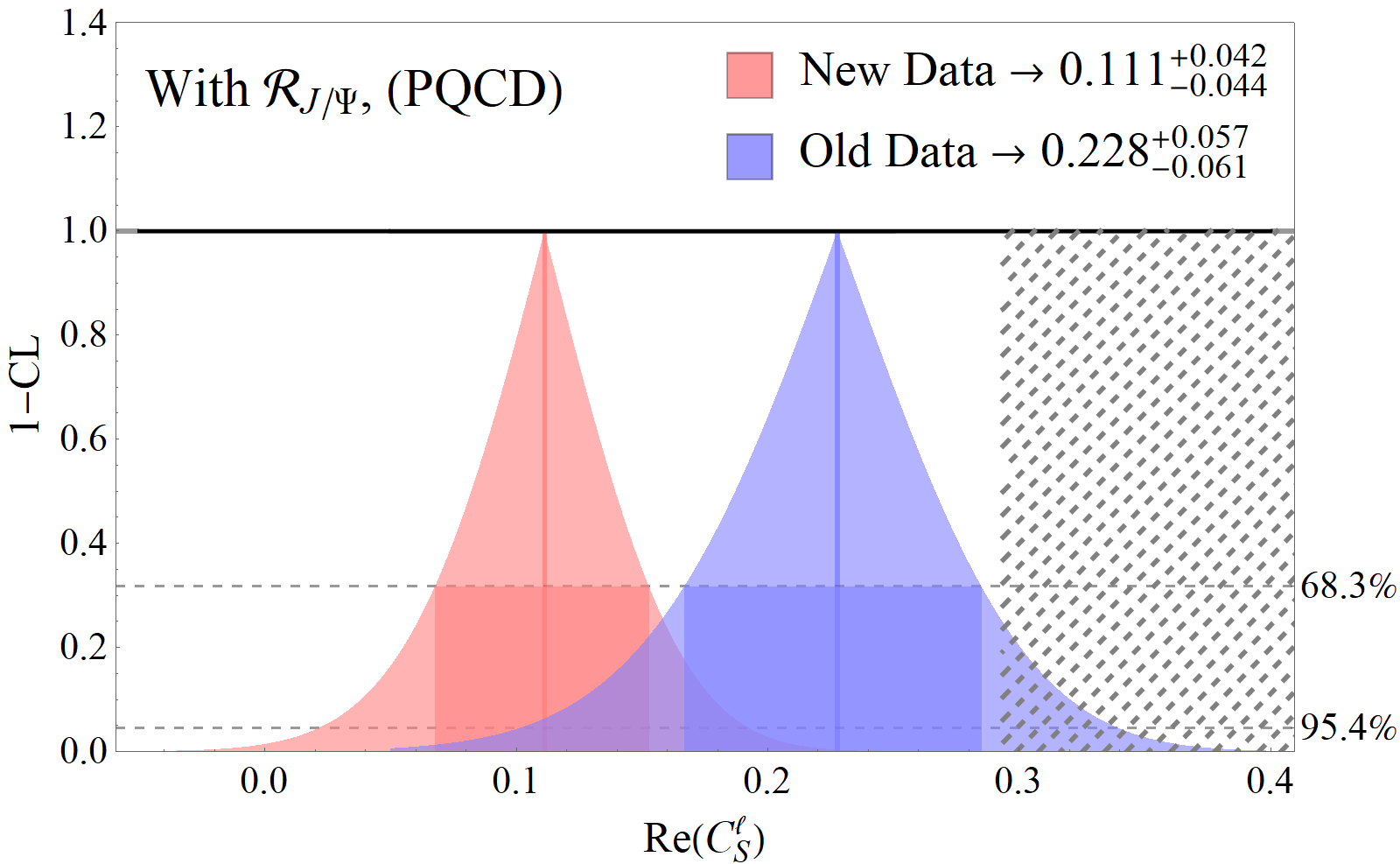}}\\
		\subfloat[LFCQ]{\includegraphics[width=0.5\linewidth]{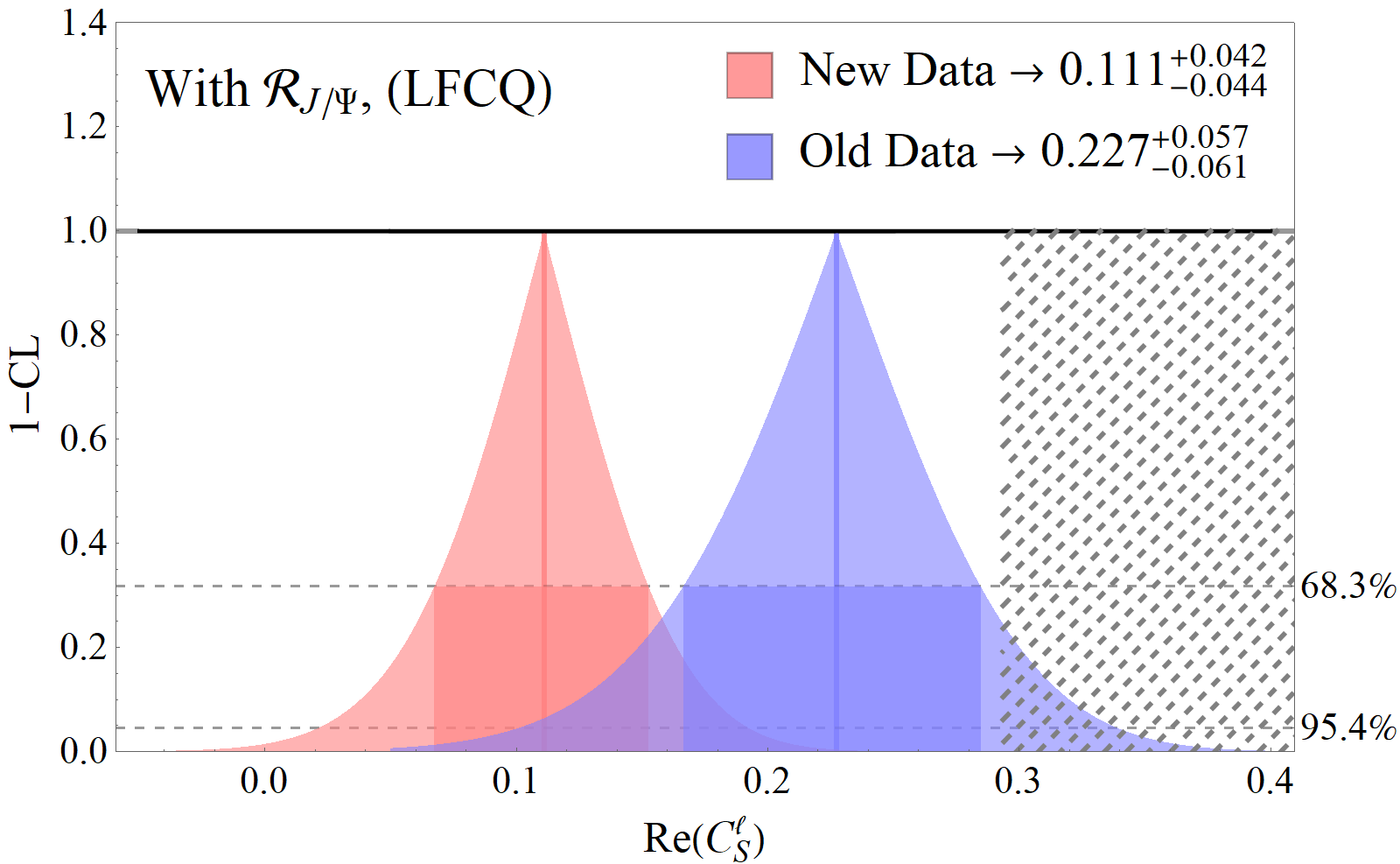}}
		\caption{One dimensional profile likelihoods, $1$ and $2\sigma$ CLs for a real $C^{\tau}_{S}$ 
		(a)without $\mathcal{R}_{J/\psi}$, (b)with $\mathcal{R}_{J/\psi}$ involving form factors parametrized by PQCD and (c)with $\mathcal{R}_{J/\psi}$ parametrized by LFCQ form factors. The diagonally hatched region is ruled out by the constraint $\mathcal{B}(B_c \to \tau \nu) \lesssim 10\%$. Profile likelihoods in red represent the new dataset while those in blue correspond to the old dataset.}
		\label{fig:rcH}
	\end{figure*}
	
	Clearly, eq.~\ref{csl2} ensures that $C_H$ is real and $\geq 0$ for the GM model. We can only get $C_H \to 0$ by making $\theta_H\to 0$ and/or $m_{H^{\pm}_3}\to \infty$. Setting Im$(C^{\tau}_{S})=0$ in our numerical analysis and re-fitting the data, we obtain fig.~\ref{fig:rcH}. We essentially get the same best-fit points for the real part of $C^{\tau}_{S}$ as in the earlier two-parameter fits (as best fit values of Im(${C^{\tau}_{S}}$) were all consistent with zero). It is evident from fig.~\ref{fig:rcH} that the values for Re($C^{\tau}_{S}$), although different for the new and the old datasets, are not affected appreciably by the inclusion or exclusion of the $\mathcal{R}_{J/\psi}$ data. One can also clearly see that the best fit value for Re($C^{\tau}_{S}$) is positive (upto $\sim 2.9\sigma$ for the new dataset and $\sim 3.7\sigma$ for the old one), which from eq.~\ref{csl2} results in a negative $C_H$. A negative value for $C_H$ in the case of the GM model is physically impossible, $C_H$ being the ratio of the squares of two real quantities. Hence a purely scalar explanation of the current $b\to c\tau\nu$ data in the framework of the GM model is severely constrained. However, the recent Belle (2019) data superseding the Belle (2016)-I data results in this tension being brought down from $\sim 3.7\sigma$ to $\sim 2.9\sigma$.
	
	\begin{figure}[hbt]
		\centering
		\includegraphics[scale=0.6]{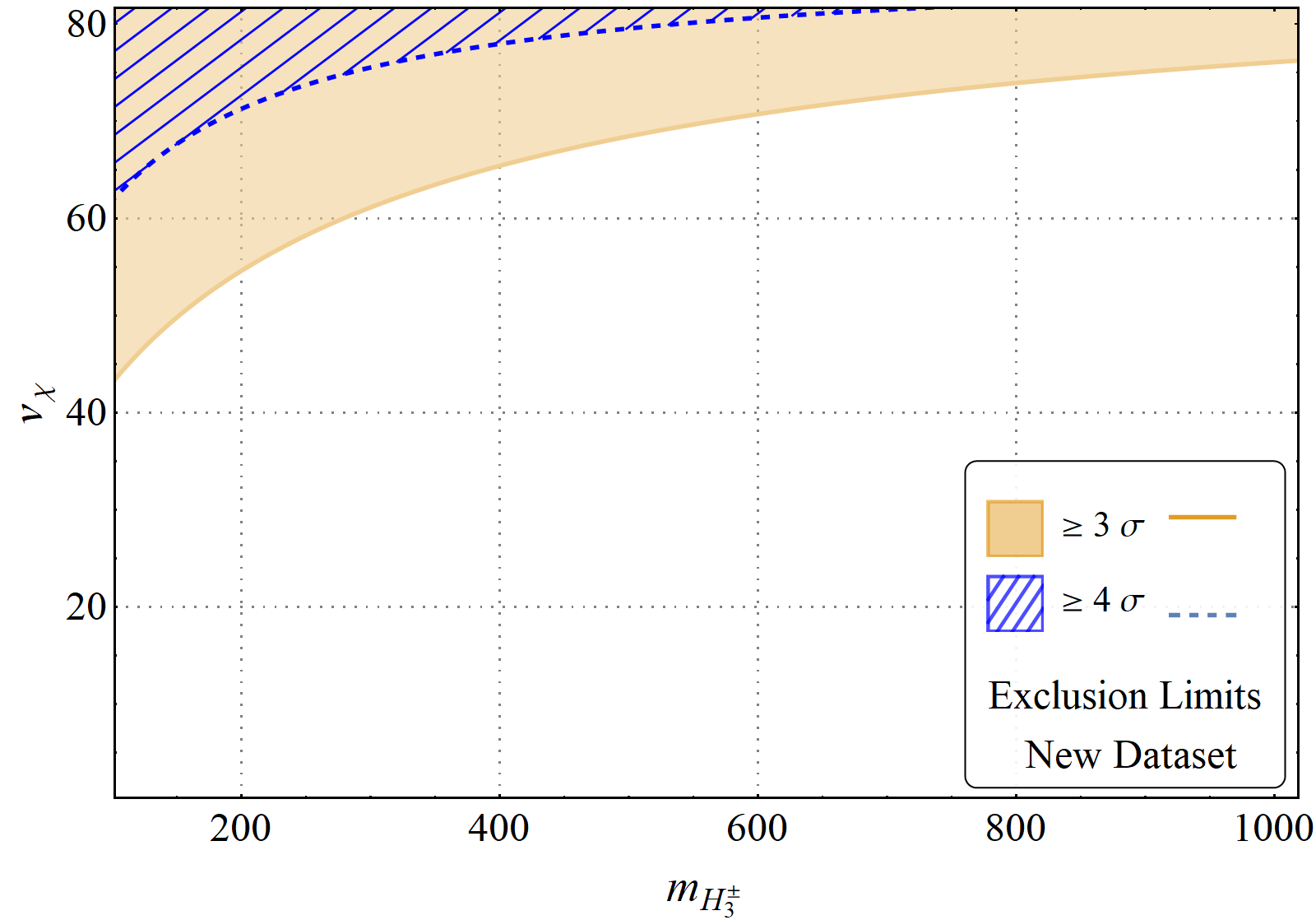}
		\caption{$v_{\chi}$ vs. $m_{H^{\pm}_3}$ parameter space excluded by all Belle and LHCb data at $2\sigma$ (orange, solid) and $3\sigma$ (blue, dashed) confidence levels. Regions above the lines are excluded.}
		\label{fig:excGM}
	\end{figure}
	
	These findings, though specific to this model, can, in general, be applicable to any model with an extended Higgs sector. For such models, contribution to $C_H$ appears to be proportional to a term like ($\tan^2 \theta_H~/~m_{H^\pm_3}^2$). As these terms are naturally $\gtrsim 0$, a preferably negative solution to $C_H$, as found in this analysis, puts a huge constraint on a purely scalar explanation of the current $b\to c\tau\nu$ data in the framework of such models.
	

\subsection{Leptoquark Model}
\subsubsection{Description}
We now turn towards an NP model that allows for the operator ${\cal O}_{S_1}$ to be preceded by a complex Wilson coefficient. While there are other such models that have been discussed in the literature, the Leptoquark (LQ) models\footnote{For a detailed review on the LQ models, the interested reader is referred to ref.~\cite{Dorsner:2016wpm}.} are particularly interesting from a phenomenological point of view owing to the diverse types of interaction (based on the Lorentz structure) that are allowed in this model. A general LQ model is comprised of twelve new particles, namely LQ particles, that carry both baryon and lepton numbers and interact with quarks as well as leptons simultaneously. Among these particles, six are scalars and the rest six are vectorial in nature under the Lorentz transformation. A considerable amount of work has been done on the LQ models - both from the perspective of flavor and of collider physics~\cite{Dorsner:2014axa,Allanach:2015ria,Evans:2015ita, Li:2016vvp, Diaz:2017lit,Dumont:2016xpj,Faroughy:2016osc,Greljo:2017vvb,Dorsner:2017ufx,Allanach:2017bta,Crivellin:2017zlb,Hiller:2017bzc,Buttazzo:2017ixm,Calibbi:2017qbu,Sahoo:2016pet, Altmannshofer:2017poe,Assad:2017iib,DiLuzio:2017vat}. Now in this model the relevant $b \to c \ell \nu$ transition is possible via 6 LQ particles\footnote{In ref. \cite{Sahoo:2016pet}, the authors have also studied ($\mathcal{R}_{D^{(*)}}$) in addition to $\mathcal{R}_{K}$. However, we address the anomalies by considering the only one non-zero Wilson coefficient $C_{S_1}^l$ and have incorporated the $\mathcal{R}_{J/\psi}$ correlation in our current article.}. 
Using the Lagrangian (introduced by Buchmuller et al.~\cite{Buchmuller:1986zs}) with the general dimensionless flavor non-diagonal couplings of scalar and vector LQs invariant under the gauge group $SU(3)_c\times SU(2)_L\times U(1)_Y$ (which satisfies baryon and lepton number conservation), we write the interaction Lagrangian relevant to $b\rightarrow cl\nu$ decays following ref. \cite{Sakaki:2013bfa}:
\bea
	\mathcal{L}^{LQ}&=&\mathcal{L}_{F=0}^{LQ}+\mathcal{L}_{F=-2}^{LQ},\nonumber\\
	\mathcal{L}_{F=0}^{LQ}&=&(h_{1L}^{ij}\bar{Q}_{iL}\gamma^\mu L_{jL}+h_{1R}^{ij}\bar{d}_{iR}\gamma^\mu\ell_{jR})U_{1\mu} 
        +h_{3L}^{ij}\bar{Q}_{iL}\boldmath\sigma\gamma^\mu L_{jL}\boldmath{U_{3\mu}}+(h_{2L}^{ij}\bar{u}_{iR}L_{jL}\nonumber\\
	&&+h_{2R}^{ij}\bar{Q}_{iL}i\sigma_2\ell_{jR})R_2,\label{Lepto_lag1}\\
	\mathcal{L}_{F=-2}^{LQ}&=&(g_{1L}^{ij}\bar{Q}_{iL}^ci\sigma_2 L_{jL}+g_{1R}^{ij}\bar{u}_{iR}^c\ell_{jR})S_1
	+g_{3L}^{ij}\bar{Q}_{iL}^ci\boldmath\sigma_2\sigma L_{jL}\boldmath{S_3}+(g_{2L}^{ij}\bar{d}_{iR}^c\gamma^\mu L_{jL}\nonumber\\
	&&+g_{2R}^{ij}\bar{Q}_{iL}^c\gamma^\mu\ell_{jR})V_{2\mu},\label{Lepto_lag2}
\eea
where $Q_i$ and $L_j$ are quark and lepton $SU(2)_L$ doublets respectively. $u_{iR}$, $d_{iR}$ and $\ell_{jR}$ are the right-handed up, down quark and charged lepton $SU(2)_L$ singlets. The $i$ and $j$ indices denote the quark and lepton generations. $\psi^c = C\bar\psi^T=C\gamma^0\psi^*$ is a charge-conjugated fermion field. The color indices are suppressed for simplicity. The quantum numbers of the LQs are provided in Table~\ref{tab:LQ_numbers}.
\begin{table*}[h!]
		\centering
                     \resizebox{8cm}{!}{
			\begin{tabular}{|c|c|c|c|c|c|c|}\hline
				& $S_1$  & $S_3$  & $V_2$  & $R_2$  & $U_1$  & $U_3$ \\
				\hline
				spin              & 0      & 0      & 1      & 0      & 1      & 1 \\
				$F=3B+L$          & -2     & -2     & -2     & 0      & 0      & 0 \\
				$SU(3)_c$         & 3$^*$  & 3$^*$  & 3$^*$  & 3      & 3      & 3 \\
				$SU(2)_L$         & 1      & 3      & 2      & 2      & 1      & 3 \\
				$U(1)_{Y=Q-T_3}$  & 1/3    & 1/3    & 5/6    & 7/6    & 2/3    & 2/3 \\ \hline
			\end{tabular}}
		\caption{Quantum numbers of scalar and vector Leptoquarks with $SU(3)_c\times SU(2)_L\times U(1)_Y$ invariant couplings.}
		\label{tab:LQ_numbers} 
\end{table*}
Here $S_{1, 3}$ and $R_2$ are the scalar LQ bosons, $U_{1, 3}^\mu$ and $V_2^\mu$ are the vector LQs. 
In this work, we investigate only the $V_2^\mu=(3, 2, 5/6)$ vector LQ which mediates the $b \to c \tau {\nu}$ quark level transitions such that except the $C_{S_1}^l$ all other Wilson coefficient are zero. The interaction Lagrangian of $V_2^\mu$ LQ with the SM fermion bilinear is given in the following:
\bea
\mathcal{L}^{LQ}&=&\left(g_{2L}^{ij}\bar{d}_{iR}^c\gamma^\mu L_{jL}+g_{2R}^{ij}\bar{Q}_{iL}^c\gamma^\mu\ell_{jR}\right)V_{2\mu}.
\label{Lagrangianv2}
\eea
The fermion fields in eq.~\ref{Lagrangianv2} are given in the gauge eigenstate basis in which the Yukawa couplings of the charged leptons and the up-type quarks are diagonal. Performing the necessary rotations (of the down-type quarks) to the mass eigenstate basis and the Fierz transformations, the Wilson coefficients relevant for the $b\rightarrow c \tau\nu$ at the LQ mass scale are given by:
\begin{equation}\label{cs_LQ}
C_{S_1}^l = { 1 \over 2\sqrt2 G_F V_{cb} } \sum_{k=1}^3 V_{k3} \left[ -{2g_{2L}^{kl}g_{2R}^{23*} \over M_{V_2^{1/3}}^2}\right].
\end{equation}
Finally it is necessary to evolve $C_{S_1}^l$ down from the scale of the LQ mass to scale of $m_b$, the mass of the bottom quark. This is achieved via the relation
\bea\label{cs_running}
C_{S_1}^l(\mu_b) &=& \left[ \alpha_s(m_t) \over \alpha_s(\mu_b) \right]^{\gamma_S \over 2\beta_0^{(5)}} \left[ \alpha_s(m_{\rm LQ}) \over \alpha_s(m_t) \right]^{\gamma_S \over 2\beta_0^{(6)}} C_{S_1}^{kl}(m_{\rm LQ}) \nonumber\\
&=&-\left[ \alpha_s(m_t) \over \alpha_s(\mu_b) \right]^{\gamma_S \over 2\beta_0^{(5)}} \left[ \alpha_s(m_{\rm LQ}) \over \alpha_s(m_t) \right]^{\gamma_S \over 2\beta_0^{(6)}} { 1 \over 2\sqrt2 G_F V_{cb} } \sum_{k=1}^3 V_{k3} \left[{2g_{2L}^{kl}g_{2R}^{23*} \over M_{V_2^{1/3}}^2}\right].
\eea
In our numerical analysis, for the sake of simplicity, we will neglect doubly Cabibbo suppressed $\mathcal{O}(\lambda^2)$ terms and keep only the leading terms proportional to $V_{33}\equiv V_{tb}$. The anomalous dimensions of $\gamma_S = 6 C_F = 8$ and $\beta_0^{f}=11-2n_f/3$~\cite{Dorsner:2013tla}.

Although most of the recent constraints from direct searches by the ATLAS and CMS collaborations are on scalar LQ masses \cite{Aad:2015caa,Aaboud:2016qeg,Khachatryan:2014ura}, there are a few references which provide updated bound on the masses of vector Leptoquarks~\cite{DiLuzio:2017vat}. Following the arguments therein, we assume that $M_{V_2^{1/3}} \approx 1.4$ TeV
in our numerical analysis. The b-quark scale is chosen to be $\mu_b = m_b = 4.2$ GeV.

\subsubsection{Analysis}
We present our numerical results for the LQ model in Table~\ref{tab:LQ_Ch}, using eqs.~\ref{cs_LQ} and \ref{cs_running}. Here we show the allowed $1 \sigma$ ranges of both real and imaginary parts of the product of the couplings involved in the $b\rightarrow c \tau\nu$ transitions via the $C_{S_1}^{\tau}$ Wilson coefficient. These are obtained from the best-fit results of Table~\ref{tab:res1} and using $V_{cb} = 39.77(89)\times 10^{-3}$~\cite{Jaiswal:2017rve}, $V_{tb}=1.009(31)$~\cite{Patrignani:2016xqp}. The nature of the allowed regions in the 2-dimensional parameter space would be similar to the regions obtained in fig.~\ref{fig:rcHicH}, although the axes will be scaled according to eq.~\ref{cs_running}. It is evident from this table that the ranges for the allowed real and imaginarily parts of the VLQ coupling product corresponding to the $\tau$ lepton do not change for the with and without $\mathcal{R}_{J/\psi}$ cases, although they span a different range for the new dataset in comparison to the old one.
\begin{table*}[h!]
		\centering
                     \resizebox{10cm}{!}{
			\begin{tabular}{|c|c|c|c|}\hline
		\multicolumn{2}{|c|}{Dataset}                                                      & ${\rm Re}\left(g^{33}_{2L}g^{23*}_{2R}\right)$     &  ${\rm Im}\left(g^{33}_{2L}g^{33*}_{2R}\right)$    \\\hline
		Old & without $\mathcal{R}_{J/\psi}$ & -0.49(13) & 0.00$\pm$1.18 \\  
		\cline{2-4}
		   & with $\mathcal{R}_{J/\psi}$ LFCQ & -0.49(13) & 0.00$\pm$1.16 \\
		\cline{2-4}
		   & with $\mathcal{R}_{J/\psi}$ PQCD & -0.49(13) & 0.00$\pm$1.15 \\
		\hline                  
		New & without $\mathcal{R}_{J/\psi}$ & -0.240(94) & 0.00$\pm$1.06 \\
		\cline{2-4}
		   & with $\mathcal{R}_{J/\psi}$ LFCQ & -0.241(94) & 0.00$\pm$1.05 \\
		\cline{2-4}
		   & with $\mathcal{R}_{J/\psi}$ PQCD & -0.241(94) & 0.00$\pm$1.04 \\
		\hline
		\end{tabular}}
		\caption{Allowed values of the product of the couplings (both real and imaginary) of the chosen Leptoquark model involved with the Wilson coefficient $C_{S_1}^l$.} 
			\label{tab:LQ_Ch}
\end{table*}
We should mention here that although we provide the values for the coupling products corresponding to the ``with $\mathcal{R}_{J/\psi}$'' for the sake of completion, the last two rows of table~\ref{tab:LQ_Ch} are redundant since these correspond to a p-value of $<5\%$ and hence are not acceptable fit results from a frequentest point of view.

\section{Summary}

Motivated by the recent and proposed searches for non-standard Higgses in colliders in one hand and repeated reporting of signatures of lepton flavor universality violation by experiments in the other, we have studied the effect of an effective scalar-type NP interaction in the $b\to c\ell\nu$ transitions and associated anomalies in this work. We have performed a model independent analysis of such an interaction and have found the parameter space of the associated general complex Wilson coefficient, allowed by present data. We have then elaborated our results with the help of two NP scenarios that render the Wilson coefficient real and complex, respectively. The GM model is taken to be an example of the former type and a specific type of vector LQ model as an example of the latter.

Analyzing the results of our analysis, we find that for a single effective scalar type NP to explain all data on the present charged current anomalies, the preceding Wilson coefficient, if real, has to be positive to yield better fits to the data than the SM. However, real scalar-type Wilson coefficients, which appear in models with extended Higgs sector are functions of ratios of two squared real quantities ($\tan^2\theta_H / m_{H^\pm_3}^2$) with an overall negative sign, which makes them negative. We hence conclude that a purely scalar explanation of the present data for charged current anomalies is disfavored for all models with extended Higgs sector at $\sim 3\sigma$.

We also find that the best-fit points do not depend on the inclusion or exclusion of the $\mathcal{R}_{J/\psi}$ data but are affected appreciably as a result of the very recent Belle (2019) data superseding the Belle (2016)-I one, as are the quality of the fits. In fact, with a p-value of $<5\%$, the ``with $\mathcal{R}_{J/\psi}$'' cases (for both the PQCD and LFCQ parametrizations) do not yield an acceptable fit from a frequentest point of view for the ``new'' dataset.

For completion and comparison, we find that models with a {\it complex Wilson coefficient (like the vector LQ model discussed in the paper) are still allowed by not only the available data, but constraints such as $\mathcal{B}(B_c\to\tau\nu)$  as well}.

\section{Acknowledgement}
	S.K.P. is supported by the grants IFA12-PH-34 and SERB/PHY/2016348. A.B., A.S., and S.K.P. thank Dr. Joydeep Chakrabortty (IIT Kanpur) for useful discussions and insight. S.K.P. thanks S. Uma Sankar for clarifying the case with $\mathcal{R}_{J/\psi}$.

\providecommand{\href}[2]{#2}\begingroup\raggedright\endgroup

 \end{document}